\documentclass{iau}
\usepackage{graphicx}

\title[Helioseismology of solar magnetic features] 
{Probing the structure of local magnetic field of solar features with helioseismology}

\author[Khalil Daiffallah]   
{Khalil Daiffallah$^1$
 }

\affiliation{$^1$Observatory of Algiers, CRAAG, \\Route de l'Observatoire, BP
  63, Bouzar\'eah 16340, Algiers, Algeria \\ email: {\tt k.daiffallah@craag.dz}}

\pubyear{2013}
\volume{302}  
\pagerange{119--126}
\jname{Magnetic Fields Throughout Stellar Evolution}
\editors{A.C. Editor, B.D. Editor \& C.E. Editor, eds.}
\begin{document}

\maketitle

\begin{abstract}
Motivated by the problem of local solar subsurface magnetic structure, we have
used numerical simulation to investigate the propagation of waves through
monolithic magnetic flux tubes of different size. A cluster model can be a good approximation to simulate sunspots as well as solar plage regions which are composed of an ensemble of compactly packed thin flux tubes. Simulations of this type is a powerful tool to probe the structure and the dynamic of various solar features which are related directly to solar magnetic field activity.
\end{abstract}

\firstsection 
\section{Introduction}
Understanding the origin of the Sun's magnetic field is the most important topic in solar physics today.
Sunspots are a manifestation of strong magnetic field  at the
surface. Other magnetic features in the form of magnetic flux tubes
can be distinguish, like a plages which are a concentration of small-scale magnetic flux tubes (bright area in solar surface) and pores which are
an isolated vertical magnetic flux tubes.
Constraining the subsurface structure, dynamics  and evolution of these
magnetic features is essential to establish a relationships
between internal solar properties and magnetic activity in the
photosphere. Helioseismology is a powerful tool to probe
the structure and dynamics of the Sun through the observation of solar
oscillations. However, the method of local helioseismology is still
limited since the magnetic
field is not included in the theory. We only interpet the observations in the
quite Sun in term  of temperature variation or velocity flow, but not in
sunspots where there is a strong magnetic field. Therefore, numerical
simulations are needed to infer the structure of the magnetic field by
modeling the interaction of waves with magnetic features.

In the first part of this study (Section \ref{part1}), we investigate the propagation of a
linear surface gravity wave packet
($f$-mode) through a monolithic structure of magnetic flux tube of different
size. In the second part (Section \ref{part2}), we explore the helioseismic response of a cluster
model which considere the subsurface magnetic field of sunspots and
plages as a bundle of small-scale magnetic flux tubes like spaghetti configuration.

\section {Helioseismic signature of monolithic magnetic flux
  tubes}
\label{part1}
We use the {\sf{SLiM}} code (\cite[Cameron et al. 2007]{cameron07}) to propagate
a linear and Gaussian $f$-mode wave packet ($\nu=3$ mHz) through a three
dimensional enhanced polytropic atmosphere (\cite[Cally \& Bogdan 1997]{cally97}).

In this section, we explore the helioseismic response of magnetic flux tubes with radii
from 200 km (e.g., internetwork magnetic field) to 3 Mm (e.g., pore or
small sunspot).  We considere 4820 G purely vertical flux tube along
$z$-direction. The flux tube is almost evacuated and it is superposed on the
background atmosphere. The scattered
wave field is constructed as the difference between the simulation with and
without the flux tube. Different scattered wave field patterns were observed for the different flux
tubes. When the flux tube is small compared to the value of $\lambda/2\pi$
where $\lambda$ is the wavelength of the incident
wave, mainly the $m=1$ kink modes are excited ($m$ is the azimutal
number of the wave). For mid-ranged tubes, the
oscillations are a mixture of $m=1$ kink mode, and $m=0$ sausage modes. For
larger tubes, numerous modes with various $m$ are excited (\cite[Daiffallah et al. 2011]{daiffallah11}).

\begin{figure}
\begin{center}
 \includegraphics[width=3.2in]{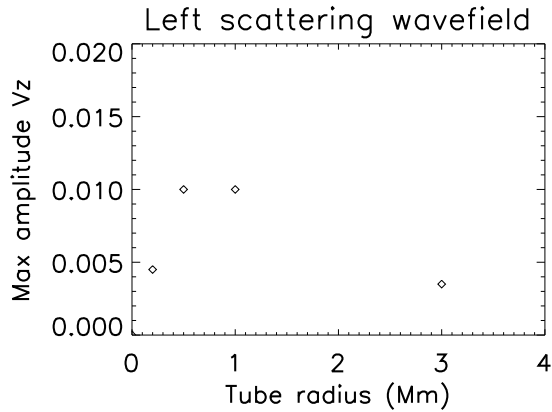} 
 \caption{A plot of the maximum scattering amplitude as a function of tube
   radius measured in the far field.}
   \label{fig1}
\end{center}
\end{figure}

If we plot the maximum scattering amplitude versus tube radius measured in the
far field (Fig.\,\ref{fig1}), we find that tubes with radii around
$\lambda/2\pi$ are the largest scatterers. We explain the decrease of
scattering  after this value of radius by the excitation of $m$-modes and the absorption
of waves by these large tubes. For example in the case of tube with $R=3$ Mm,
a part of $f$-mode is converted to slow magneto-acoustic-gravity mode which
propagates along magnetic field lines in $z$-direction (e.g., \cite[Cally 2005]{cally05}). This process can explain the observations of \cite[Braun et al.(1987)]{braun87} and 
 \cite[Braun et al.(1988)]{braun88} which reveal a deficient in the power of 
the outgoing wave compared to the incoming wave from a typical sunspot on the solar surface (Fourier-Hankel analysis).

\begin{figure}
\begin{center}
 \includegraphics[width=2.6in]{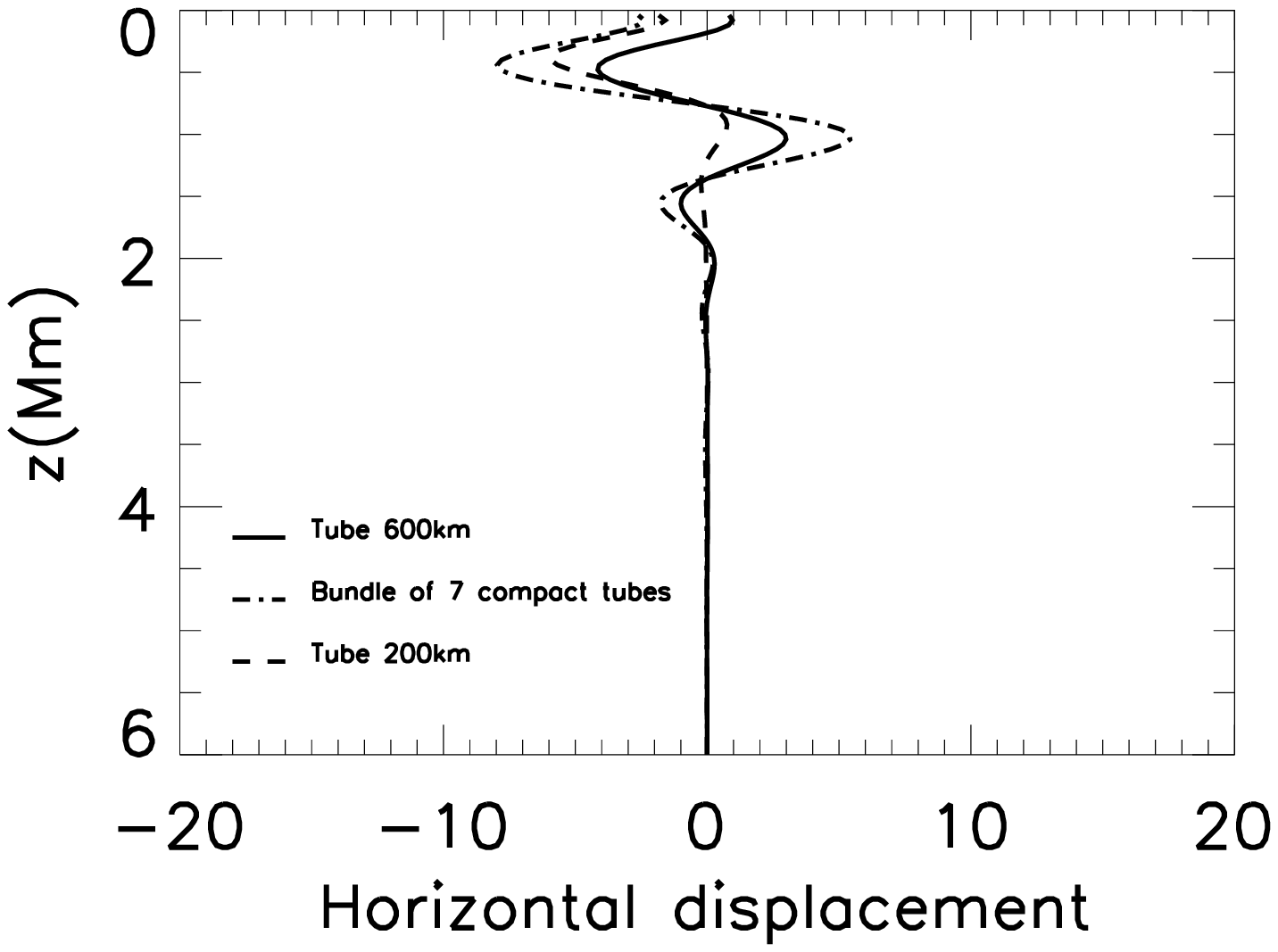} 
 \includegraphics[width=2.6in]{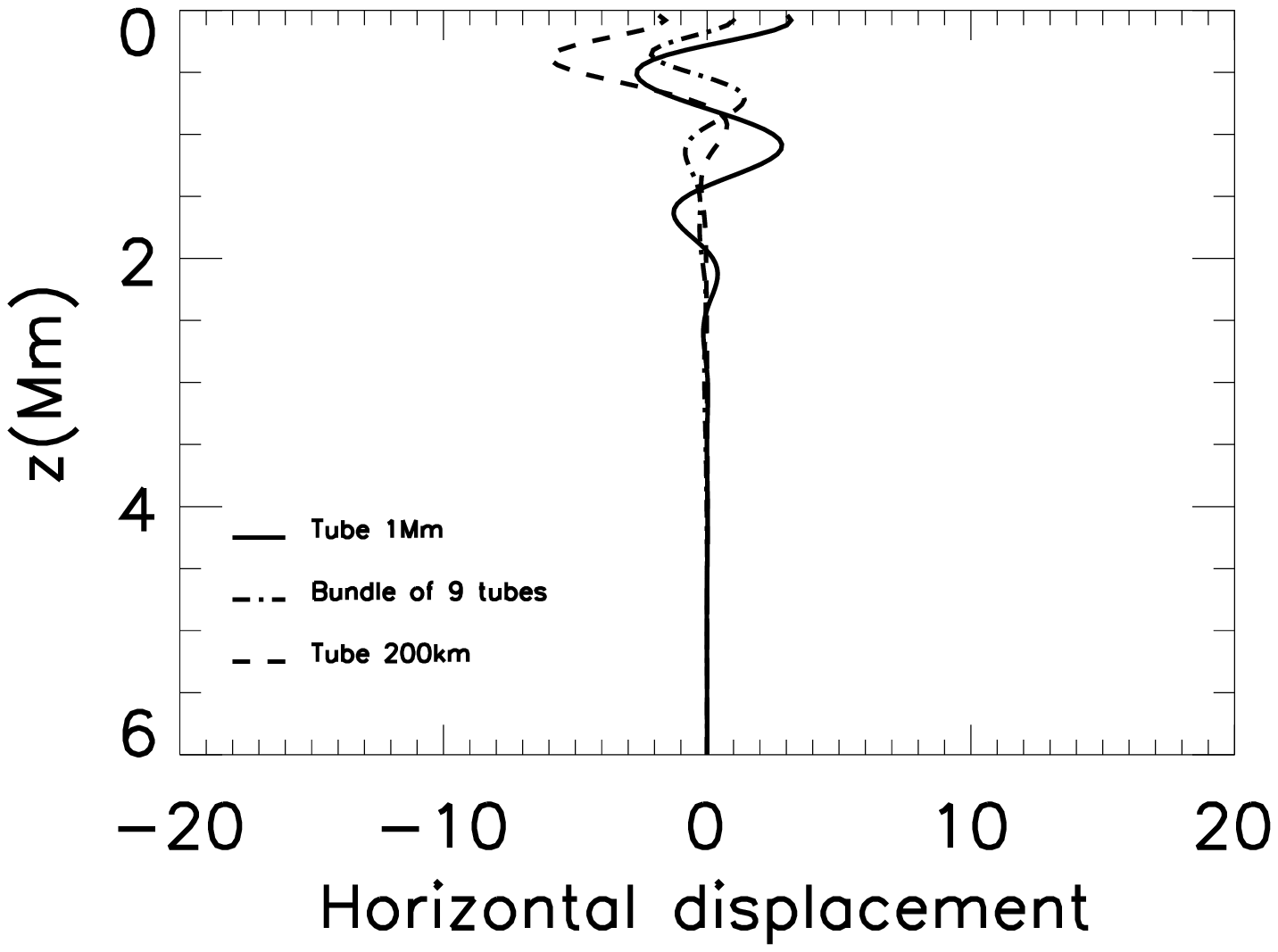} 
 \caption{ Horizontal displacement of the central tube axis of the compact cluster in the left panel, and the loose cluster in the right panel (dot-dashed line), as a function of depth $z$. 
The dashed and the solid lines show the oscillations of a single tube of 200 km radius and the monolithic equivalent tube, respectively. The separation
between the tubes inside the loose cluster is about $\lambda/2\pi$. }
   \label{fig2}
\end{center}
\end{figure}

\section {Helioseismic response of a cluster of small magnetic flux tubes}
\label{part2}

In this section,  we want to know what is the structure of magnetic field beneath
sunspots: is it like a monolithic model as in section \ref{part1} or  like a
cluster model? how to distinguish from the observed oscillations between these two models? 
The cluster model can be a good approximation to simulate solar plage regions which are
composed of an ensemble of compactly packed thin flux tubes (\cite[Hanasoge \& Cally 2009]{cally09}).

Motiveted by these problems, we investigate the
propagation of waves ($f$-mode) through a cluster of small identical magnetic flux tubes of 200 km
radius (e.g., \cite[Hindman 
\& Jain 2012]{hindman12}; \cite[Daiffallah 2013]{daiffallah13}; \cite[Felipe et al. 2013]{felipe13}).

We have studied two cases, one is a compact cluster which it consists of seven
identical magnetic flux tubes in a hexagonal close-packed configuration, and
the other case is a loose cluster of nine tubes.

\begin{figure}
\begin{center} 
 \includegraphics[width=2.7in]{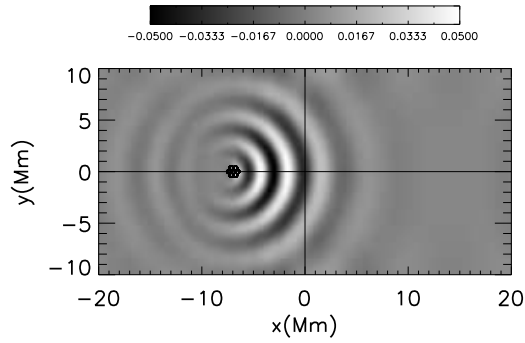} 
 \caption{A snapshot at $t$ = 3300 s of the scattered wave field (Vz) of a compact 
   cluster of seven identical tubes of 200 km radius.}
   \label{fig3}
\end{center}
\end{figure}

The Fig.\,\ref{fig2} shows the horizontal displacement of the central tube axis as a function of depth $z$ (dot-dashed line) for the compact cluster in left panel, and the loose cluster in right panel. We can observe that the upper part of 
the compact cluster oscillates more like a single tube of 200 km radius (dashed line) than like the monolithic equivalent tube of 600 km (solid line). Furthermore, the large amplitude of the compact cluster compared to the amplitude of 200 km single tube confirm that this oscillation concerne the whole part of the compact cluster and not the tubes individually. This can be seen in the scattered wave field of the compact cluster in the Fig.\,\ref{fig3} where the compact cluster seems oscillate like a single object and there is no observation of multiple scattering in the near field. 

The amplitude of the loose cluster in the right panel of Fig.\,\ref{fig2} is smaller than the amplitude of the 200 km single tube,  which means that the incident wave energy is converted to tubes oscillation and the oscillation of the loose cluster in Fig.\,\ref{fig2} corresponds to that of the individual tubes. Therefore, the loose cluster will show multiple scattering from the individual tubes in the near field. In this case, the absorption of the incident wave by the cluster will be enhanced.

\section{Discussion}
Sunspots are a manifestations of strong magnetic field at the solar surface. They represent a major relation between internal magnetic
field and solar activity in the photosphere. 
Local helioseismology is a powerful tool to investigate the 
substructure of the Sun. However, interpretation of data have been somewhat
ambiguous in solar active regions where the magnetic field is strong. Numerical simulations provide an efficient and direct way to understand the helioseismic signature of solar magnetic features, which have recently begun to be observed. We simulate in this study the propagation of a linear $f$-mode wave packet through different magnetic features. The aim is to get informations about the characteristics and the structure of these features by studying the scattered waves observed at the solar surface.  The principal results of the simulations can be summarized in the following way:

\begin{itemize}
\item Magnetic flux tubes are a strong absorbers and  scatterers  of  $f$-mode waves.
\item Different scattered wave field patterns were observed for different monolithic magnetic flux tubes radii. This will allow us to inferring the typical size of magnetic structures, but also magnetic field strength, vector orientation,  profile, ...
\item  A cluster model of magnetic flux tubes (compact or loose) scatters waves in different way from a monolithic model. This will allow us to inferring the magnetic structure of more complex solar features (sunspots, plages, ...) and distinguish between monolithic or cluster model,  compact or loose cluster.

\item Determining the parameters of these magnetic features, that is, structure, typical size,  field strength, ... will help to reveal details of the process of solar dynamo and how magnetic field is transported up through the convection zone. 
\item The subsurface structure of sunspots is still poorly understood, we need (i) high-resolution space observations, (ii) improving simulations.

\end{itemize}

\end{document}